\documentclass{article}
\usepackage{graphicx}
\usepackage[utf8]{inputenc}
\usepackage{amsmath, amssymb}

\title{Spectral distortions in CMB by the bulk Comptonization due to Zeldovich pancakes}

\author{G.S. Bisnovatyi-Kogan \thanks{Space Research Institute, Profsoyusnaya 84/32, Moscow, Russia 117997; National Research Nuclear University MEPhI, Kashira Highway, 31, Moscow,
  115409; and Moscow Institute of Physics and Technology MIPT, Institutskiy Pereulok, 9, Dolgoprudny, Moscow region, 141701.}}

\date{}

\begin{document}

\maketitle
\begin{abstract}
If the large scale structure of the Universe was created, even partially, via Zeldovich pancakes, than the fluctuations of the CMB radiation should be formed due to bulk comptonization of black body spectrum on the contracting pancake. Approximate formulaes for the CMB energy spectrum after bulk comptonization are obtained. The difference between comptonized energy spectra of the CMB due to thermal and bulk comptonozation may be estimated by comparison of the plots for the spectra in these two cases.
\end{abstract}

\section{Introduction}

Fluctuations of CMB are created mainly by primordial quantum perturbations, growing due to inflation \cite{peeb93}. Also they may appear as a result of the interaction of CMB with a hot plasma
appearing during the epoch of the formation of a large scale structure of the universe. Perturbations of CMB spectrum created by this interaction had been investigated in \cite{wey1}, \cite{wey2},
\cite{peeb70}, \cite{sper17}, and in the series of papers \cite{ZS1} - \cite{ZS4}. The review of theory of interaction of free electrons with a radiation is presented in \cite{rew}.

A main process of interaction of hot plasma inside galaxy clusters with CMB is a Compton scattering, including a Doppler effect. Observations from the satellites WMAP and PLANCK had revealed
an existence of CMB fluctuations in the directions of rich galactic clusters \cite{w1}, \cite{w2}, \cite{p8}-\cite{p}, which had been interpreted in the frame of physical
processes, investigated in \cite{wey1}, \cite{wey2}, \cite{ZS1}, \cite{ZS2}.

  A study of interaction is based on the solution of the Kompaneets equation \cite{Komp}, see also \cite{wey1}, which is an approximate form of the radiative transfer equation
with non-coherent scattering, when energy exchange $\Delta E_{e\gamma}$ between the electron and the photon in one scattering is much less than the photon energy $E_{\gamma}$, and  kinetic energy (electron energy),
$kT_e$

\begin{equation}
\frac{\Delta E_{e\gamma}}{E_{\gamma}}=\frac{4kT_e-E_{\gamma}}{m_ec^2}\ll 1.
\label{i1}
\end{equation}
At such conditions, characteristic for the Thompson-Compton scattering in a non-relativistic plasma, the scattering term is reduced to the differential form \cite{Komp}, \cite{wey1}. The relative
value of spectral distortions induced by such scattering, is preserved during the universe expansion, because the perturbed and background photons have the same dependence on time and on the red shift,
when moving without interactions.

In addition to the scattering on hot electrons, another type of distortion of the photon spectrum takes place in the interaction of a photons with a bulk motion of a matter.
This types of distortion had been investigated in application to the accretion of matter into neutron stars, stellar mass and AGN black holes, which are observed as X-ray, optical,
and ultraviolet sources \cite{BP1}, \cite{BP2},
\cite{TZ}, \cite{PL}, \cite{PL2}, \cite{T1}, \cite{BW}, \cite{CGF}, \cite{BW2}. Among the objects, in which a bulk comptonizanion may be important, we should include Zeldovich pancakes
which may be formed in the universe at first stages of a large-scale structure formation.  \cite{ZPan}, \cite{Z2}, \cite{ZDSh}.
The  equation, including both thermal and bulk motion components, have been derived  in \cite{BP1}, \cite{PL} in different approximations.
In this work we calculate distortions of the equilibrium Planck spectrum of CMB, induced by the bulk comptonization on the objects, collapsing to Zeldovich pancakes.
We use an approximation of a uniform flat collapsing layer, which reproduces a motion of a non-spherical large-scale perturbation, with a gravity of a dark matter (DM) \cite{BK}.
We suggest, that hydrogen baryonic matter, collapsing in the gravitational field of DM, is sufficiently ionised, to be able to produce a bulk comptonization.

Observations of radio spectra in the region of the redshifted famous hydrogen radio line with a wavelength $\lambda\, =\,21$ cm, have lead to the detection of a flattened
absorption profile in the sky-averaged radio spectrum, which is
centred at a frequency of 78 megahertz, corresponding to the redshift $z\approx 17$ \cite{BNat}.
The appearance of this profile was attributed to an action of the radiation of first stars \cite{BNat,Funiv}, which are responsible also for a secondary heating of the universe.
Spectral observations of a distant quasar J1342 + 0928 with $z=7.54$ have shown  strong
evidence of absorption of the spectrum of the quasar redwards of
the Lyman $\alpha$ emission line (the Gunn–Peterson damping wing), as
would be expected if a significant amount (more than 10 per cent) of
the hydrogen in the intergalactic medium surrounding this quasar
is neutral \cite{BanNat}. The authors derive such a significant fraction of neutral hydrogen,
although the exact fraction depends on the modelling. However,
even in the most conservative analysis it was found a fraction the neutral hydrogen of more
than 0.33 (0.11) at 68 per cent (95 per cent) probability, indicating
that this redshift is  well within the reionization epoch of the
Universe.
So we suggest that formation the main body of a large-scale structure  in the model of Zeldovich pancakes happens after the period of the secondary ionization \cite{ts}, \cite{tsb}  in the universe.

 We consider a contractive  flat layer of plasma surrounded by the radiation with equilibrium Planckian spectrum (CMB). When crossing the contractive layer the photons experience a Compton scattering
 on electrons, which velocities have a thermal (chaotic), and directed (bulk motion) components. For sufficiently low-temperature plasma the bulk motion comptonization becomes more important than the
 thermal one. In Section 2 we consider bulk motion comptonization by the "cold" contractive layer. We solve analytically the Kompaneets equation in a contractive layer,
similar to the one for  converging flow in   \cite{BP1}, \cite{BW}, which is illuminated by an equilibrium radiation flux on its boundary.
In the Section 3 we compare the spectra of a thermal, and bulk motion comptonization at parameters, when both distortions are quantitatively comparable, but have different spectral shapes.

\section{The bulk comptonization by the contractive self-gravitating layer}

\subsection{Dynamics of the contractive self-gravitating layer}

There are several physical input parameters in the problem of calculation of resulting spectrum of radiation passing the flat contractive layer of plasma. We consider one-dimensional problem, when all
function depend only on one space coordinate $x$. The temperature $T$ and density $\rho$ inside the layer are supposed to be uniform, depending only on time $t$. For adiabatic contraction the thermal state
of the matter is characterized by a constant entropy $S$. For adiabatic contraction of the layer of non-relativistic ideal fully ionized plasma the equation of state, connecting pressure $P$ with density is written as

\begin{equation}
\label{eos}
P=\rho{\cal R}T=K(S)\rho^{5/3}.
\end{equation}
The pressure $P_0$ and the density $\rho_0$ in the layer at the initial moment $t_0$ are given, determining the constant $K$, and the entropy $S(K)$. For any time dependent $\rho(t)$ we obtain $P(t,S)$ from
(\ref{eos}), and for a known gas constant ${\cal R}$ we obtain a time dependent temperature

\begin{equation}
\label{T}
T(t,S)=\frac{P}{\rho{\cal R}}.
\end{equation}
We consider a self-gravitating layer with initial thickness $x_0$, initial density $\rho_0$, and initial velocity distribution over the layer as

\begin{eqnarray}
\upsilon_0(y) = - \upsilon_0\frac{y}{x_{0}}, \quad -\frac{x_0}{2}\le y \le \frac{x_0}{2},\nonumber \\
-\frac{\upsilon_0}{2}\le \upsilon_0(y)\le \frac{\upsilon_0}{2},\quad
-\upsilon_0\left(\frac{x_0}{2}\right)=\upsilon_0\left(-\frac{x_0}{2}\right)=\frac{\upsilon_0}{2}.
\label{v0}
\end{eqnarray}
Here $\upsilon_0$ is the initial velocity of decreasing of the layer thickness $\frac{dx}{dt}(0)=-\upsilon_0$.  The surface density of the layer $\sigma_0=\rho_0 x_0=\rho(t) x(t)=\sigma$  remains constant during contraction.
The time dependent parameters of the cold $(S=K=0)$ uniform layer are obtained from solution of equations of motion and Poisson, for the gravitational potential $\varphi_g(y,t)$ and he gravitational force $F_{gy}$, and
the equation for a time dependence
of the thickness of the layer $x(t)$

\begin{eqnarray}
\frac{\partial^2\varphi_g}{{\partial y}^2} = 4\pi G\rho, \,\, F_{gy}=-\frac{\partial\varphi_g}{\partial y}=-4\pi G\rho y,\,\,
F_{gy}\left(\frac{x}{2}\right)=-2\pi G\rho x=-2\pi G\sigma_0. \label{pm} \\
\frac{\partial\upsilon_y}{\partial t}=F_{gy}=-4\pi G\rho y,\,\,\,
\upsilon_y\left(\frac{x}{2}\right)=-2\pi G\rho xt-\frac{v_0}{2}=-2\pi G\sigma t-\frac{v_0}{2}.
\label{pm1}
\end{eqnarray}
Solving these equations with boundary and initial conditions, mentioned above, we obtain the solution, taking $ t_0=0$, as

\begin{eqnarray}
\upsilon(t,y) =( - \upsilon_0-4\pi G\sigma_0 t)\frac{y}{x}, \,\,\,
\frac{dx}{dt}=-2\upsilon_y\left(\frac{x}{2}\right)= - \upsilon_0-4\pi G\sigma_0 t,\nonumber\\
\quad x=x_0- \upsilon_0 t -2\pi G\sigma_0 t^2 \quad {\rm at} \quad
-\frac{x}{2}\le y \le \frac{x}{2}.
\label{v}
\end{eqnarray}
The thickness of the layer becomes $x=0$ at $t=t_1$, with

\begin{equation}
t_1=-\frac{\upsilon_0}{4\pi G\sigma_0} + \sqrt{ \frac{\upsilon_0^2}{(4\pi G\sigma_0)^2}+\frac{x_{0}}{2\pi G\sigma_0}}.
\label{th}
\end{equation}

\begin{figure}
\centering
\includegraphics[scale=0.5]{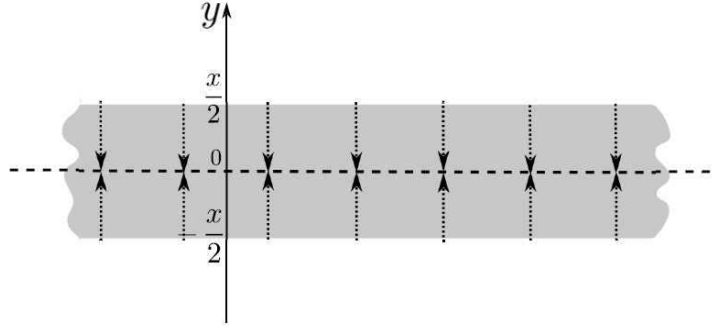}
\caption{The layer collapsing to its midplane; $y$ is a variable across the layer,
$x$ is a layer thickness.}
\label{fig:layer}
\end{figure}

 The bulk comptonization is more important  at large contraction velocity. We consider only last stages, when on the surface $\upsilon$ remains close to $\upsilon_0$. It happens when
 $\upsilon_0^2\gg 8\pi G\sigma_0 x_0$. In this approximation we obtain the following formulae describing the dynamics of the cold layer

 \begin{equation}
t_1=\frac{x_0}{\upsilon_0},\quad \upsilon(y) = - \upsilon_0\frac{y}{x}, \quad x=x_0- \upsilon_0 t \quad {\rm at} \quad
-\frac{x}{2}\le y \le \frac{x}{2}.
\label{th1}
\end{equation}

\begin{figure}
\center{\includegraphics[width=0.9\linewidth]{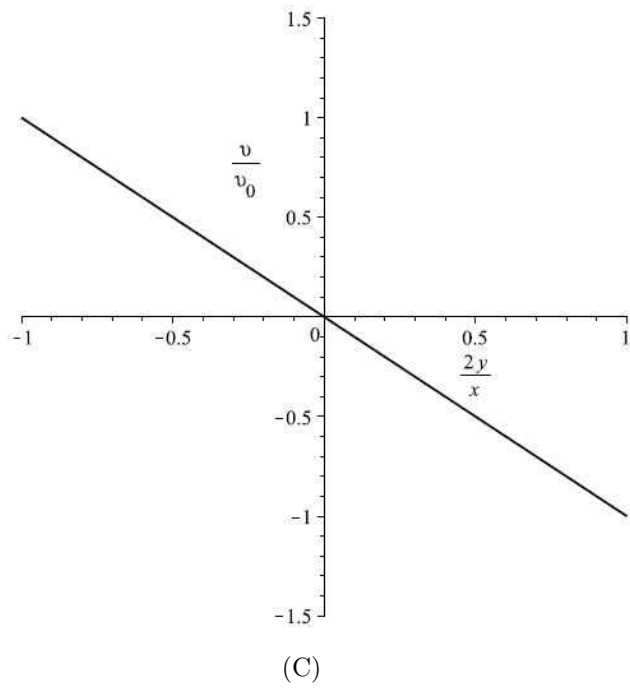} \\ (C)}
\hfill
\caption{Velocity profile in the collapsing self-gravitating plasmic layer.}
\label{fig:vel}
\end{figure}

\subsection{Equations determining the spectrum distortion due to  bulk  comptonization in the cold layer}

We consider the problem of a bulk Comptonization of an incident radiation on a moving medium in diffusion approximation using Eq.(31) of \cite{BW} (see also \cite{BP1}). The scatterer is a flat  contracting layer of cold electrons ($T_e \approx 0$), see Fig.\ref{fig:layer}. The layer is consisting mainly of a non-collisional dark matter, where particles of two oppositely moving flows penetrate through each other, imitating an elastic bounce with respect to the behaviour of the gravitational potential. The flows with baryon and electrons collide, forming a shock wave with decreasing expanding velocities \cite{ZDSh}. Therefore, the bulk comptonization is effective only on the contracting stage of the pancake formation.
Partial differential equation satisfied for the photon occupation number function $f(x,\varepsilon)$ is written as, (see \cite{BW,TMK97})

\begin{equation}
\label{eq:sol1}
\frac{\partial f}{\partial t} + \overrightarrow{\upsilon} \nabla f = \frac{\partial}{\partial y} \left(\frac{c}{3n_e\sigma_T} \frac{\partial f}{\partial y} \right) + \frac{1}{3} \left(\nabla \overrightarrow{\upsilon} \right)\nu\frac{\partial f}{\partial \nu}+j(y, \nu),
\end{equation}
where $\nu$ is a photon frequency, $n_e$ is the electron density, $\sigma_T$ is the Thompson cross-section, $j(y, \nu)$ is the emisivity.  Let us consider the respond of the system on the incoming stationary flux of monochromatic photons through both sides of the layer's boundary, described by the Green function of the problem $f_G$. Introduce following non-dimensional variables

$$
\tilde{y} = n_e\sigma_Ty, \quad \tau = \int_{-x/2}^{x/2}n_e \sigma_Tdy = n_e\sigma_Tx,$$
\begin{equation}
\label{eq:dless}
\tilde{\nu} = \frac{h\nu}{k_BT_r}, \quad \beta_0 = \frac{\upsilon_0}{c},
\end{equation}
 where $k_B$ is the Boltzmann constant, $T_r$ is the temperature of the incoming radiation, $c$ is velocity of light, $\tau$ is the constant optical depth of the contracting layer. Assuming the baryonic mass of the forming pancake at redshift $z\sim 10$ as $5\cdot 10^{14}\, M_\odot\,\approx\, 10^{48}\,g$, concentrated mainly inside the central radius $R_c\sim 0.1$ Mpc, we obtain its vertical optical depth as

\begin{equation}
\label{eq:est2}
\tau_v =  \frac{M\sigma_T}{m_p\pi R_c^2}\approx 1.5
\end{equation}
for fully ionized pancake plasma. This value of the optical depth is calculated for vertically falling background photons. For the photons falling with an inclination to pancake vertical the optical depth is larger. Therefore, qualitatively the solution for opaque pancake may be used for estimation of the bulk Comptonization effect at lower optical depths $\tau_v$.
We consider the case, when the photon birth inside the layer is negligibly small  ${j}({y},{\nu}) \approx 0$, and only scattering of the photons coming from the outer space is important.
While the contracting velocity during the pancake formation is much smaller than the light velocity, and the vertical optical depth is of the order of unity, we may consider a time independent problem for a stationary contracting velocity distribution (\ref{v0}), at fixed $\upsilon_0,\,\,\, x_0$.
In the stationary one-dimensional case the equation (\ref{eq:sol1}) is written in the form
\begin{equation}
\label{eq:sol2}
\frac{1}{3}\frac{\partial^2 f}{\partial\tilde{y}^2} + \frac{\beta_0\tilde{y}}{\tau}\frac{\partial f}{\partial \tilde{y}} - \frac{\beta_0}{3\tau} \tilde{\nu}\frac{\partial f}{\partial \tilde{\nu}} =0,
\end{equation}

\subsection{Approximate solution}

To solve the partial differential equation (\ref{eq:sol2}) approximately, we consider conditions, when perturbations of the CMB spectrum, produced by bulk comptonization are small, may represent the solution, similar to \cite{ZS1}, in the form

\begin{equation}
\label{bc1}
f=C_f(f_0+f_1),\quad f_1 \ll f_0,\,\,\, C_f \,\,\,{\rm is\,\,\, constant},
\end{equation}
 where $f_0$ is a spatially uniform planck distribution

\begin{equation}
\label{bc2}
f_0(\tilde\nu) =\frac{1}{e^{\tilde\nu}-1}
\end{equation}
Using these relations in (\ref{eq:sol2}), we obtain

\begin{equation}
\label{bc3}
\frac{1}{3}\frac{\partial^2 f_1}{\partial\tilde{y}^2} + \frac{\beta_0\tilde{y}}{\tau}\frac{\partial f_1}{\partial \tilde{y}} - \frac{\beta_0}{3\tau} \tilde{\nu}\frac{\partial f_0}{\partial \tilde{\nu}} =0,
\end{equation}
The reduce the partial differential equation (\ref{eq:sol2}) to the ordinary equation, we approximate the space dependence of $f_1$ by profiling function

\begin{equation}
\label{bc4}
f_1(\tilde\nu,\tilde y)=\tilde{f_1}(\tilde\nu)\frac{4\tilde y^2-\tau^2}{\tau^2}
\end{equation}
This function is symmetric relative to the plane $\tilde y$, and equal to zero on both boundaries of the layer, in accordance with boundary conditions. We have than

\begin{equation}
\label{bc5}
\frac{\partial f_1}{\partial \tilde{y}}=\tilde{f_1}\frac{8\tilde{y}}{\tau^2},\quad
\frac{\partial^2 f_1}{\partial \tilde{y}^2}=\tilde{f_1}\frac{8}{\tau^2}, \qquad
\frac{\partial f_0}{\partial\tilde{\nu}}=-\frac{e^{\tilde\nu}}{(e^{\tilde\nu}-1)^2}.
\end{equation}
Substituting (\ref{bc5}) into (\ref{bc3}), we obtain

\begin{equation}
\label{bc6}
\frac{1}{3}\frac{8}{\tau^2}\tilde{f_1}+
\frac{\beta_0}{\tau}\frac{8\tilde{y}^2}{\tau^2}\tilde{f_1}+
 \frac{\beta_0}{3\tau}\frac{\tilde{\nu}e^{\tilde\nu}}{(e^{\tilde\nu}-1)^2}=0.
\end{equation}
Averaging (\ref{bc6}) over the layer by integration $\int_{-\tau/2}^{\tau/2}d\tilde y$, we obtain

\begin{equation}
\label{bc7}
\frac{1}{3}\frac{8}{\tau}\tilde{f_1}+
\frac{2\beta_0}{3}\tilde{f_1}+
 \frac{\beta_0}{3}\frac{\tilde{\nu}e^{\tilde\nu}}{(e^{\tilde\nu}-1)^2}=0.
\end{equation}
We obtain from (\ref{bc7})
\begin{equation}
\label{bc8}
\tilde{f_1}=-\frac{1}{8}
\frac{\beta_0\tau}{1+\frac{\beta_0\tau}{4}}\,
 \frac{\tilde{\nu}e^{\tilde\nu}}{(e^{\tilde\nu}-1)^2}.
\end{equation}
Averaging (\ref{bc6})  by the same integration we obtain finally

\begin{equation}
\label{bc9}
f_1(\tilde\nu)=-\frac{2}{3}\tilde{f_1}=\frac{1}{12}\frac{\beta_0\tau}{1+\frac{\beta_0\tau}{4}}\,
 \frac{\tilde{\nu}e^{\tilde\nu}}{(e^{\tilde\nu}-1)^2}.
\end{equation}
To find the resulting spectrum after bulk comptonization process we should take into account, that
in this process the number of photon is conserved, while for the resulting function $f_0+f_1$ the photon number density is larger than for the background function $f_0$. We should therefore consider the resulting function in the form

\begin{equation}
\label{bc10}
f(\tilde\nu)=C_f(f_0(\tilde\nu)+f_1(\tilde\nu)).
\end{equation}
The number of photons does not change during the comptonization, what is represented by the relation

\begin{equation}
\label{bc11}
\int_0^\infty \tilde\nu^2 f(\tilde\nu)d\tilde\nu=\int_0^\infty \tilde\nu^2 f_0(\tilde\nu)d\tilde\nu.
\end{equation}
From (\ref{bc10})(\ref{bc11}) we find the constant $C_f$, what uniquely define the photon distribution function after bulk comptonization

\begin{eqnarray}
\label{bc12}
C_f=\frac{\int_0^\infty \tilde\nu^2 f_0(\tilde\nu)d\tilde\nu}
{\int_0^\infty \tilde\nu^2 f_0(\tilde\nu)d\tilde\nu+\int_0^\infty \tilde\nu^2 f_1(\tilde\nu)d\tilde\nu},\\ \quad  f(\tilde\nu)= \frac{C_f}{e^{\tilde\nu}-1}\left[1+\frac{1}{12}\frac{\beta_0\tau}{1+\frac{\beta_0\tau}{4}}\,
 \frac{\tilde{\nu}e^{\tilde\nu}}{e^{\tilde\nu}-1}\right].
\label{bc13}
\end{eqnarray}
The influence of the bulk comptonization on the energy spectrum of CMB photons may be estimated by comparison of the distorted energy spectrum $f_{Eb}(\tilde\nu)$ with the planck spectrum $f_{E0}(\tilde\nu)$, defined as

\begin{eqnarray}
\label{bc14}
  f_{Eb}(\tilde\nu)= \frac{C_f \tilde\nu^3 }{e^{\tilde\nu}-1}\left[1+\frac{1}{12}\frac{\beta_0\tau}{1+\frac{\beta_0\tau}{4}}\,
 \frac{\tilde{\nu}e^{\tilde\nu}}{e^{\tilde\nu}-1}\right],\quad f_{E0}(\tilde\nu)= \frac{\tilde\nu^3 }{e^{\tilde\nu}-1}.
\end{eqnarray}

\section{Comparison with thermal comptonization effects}

It was obtained in \cite{ZS1} that the relative distortion of the comptonized spectrum of CMB due to interaction with hot electrons is determined by the expression

\begin{equation}
\label{bc15}
\delta_{Th}= y\tilde{\nu}\frac{e^{\tilde\nu}}
{e^{\tilde\nu}-1}\left[\frac{\tilde\nu}{\tanh(\tilde\nu/2)}-4\right].
\end{equation}
Here the value $y$, for the flat universe with $\Omega=1$, is determined as \cite{ZS1}

\begin{equation}
\label{bc16}
y=n_{e0}\sigma_T c H_0^{-1}\int_0^{z_{max}}\frac{kT_e(z)}{m_e c^2}\sqrt{1+z}\,dz.
\end{equation}
Equations (\ref{bc15}),(\ref{bc16}) are derived for the uniform expanding universe, with a present hot electron density $n_{e0}$, which had been heated at redshift $z_{max}$, and have a temperature dependence $T_e(z)$. When applying this formula for the CMB comptonization by hot gas in the galactic clusters,
we may approximately take the value of $y$ in the form

\begin{equation}
\label{bc17}
y\approx n_e \sigma_T l \frac{kT_e}{m_e c^2} = \tau_{CL}\, \frac{kT_e}{m_e c^2},
\end{equation}
where $n_e$, $T_e$, $l$, $\tau_{CL}$ are the number density, temperature, size, and the optical depth of hot electrons in the galactic cluster, respectively. As a result of thermal comptonization, the photon distribution function is written as

\begin{equation}
\label{bc18}
f_{Th}(\tilde\nu)=f_0(\tilde\nu)(1+\delta_{Th})=\frac{1}{e^{\tilde\nu}-1}
\left\{1+ y \tilde\nu\frac{e^{\tilde\nu}}{e^{\tilde\nu}-1}
\left[\frac{\tilde\nu}{\tanh(\tilde\nu/2)}-4\right]
\right\}.
\end{equation}
The distorted energy spectrum is written as

\begin{equation}
\label{bc19}
f_{ETh}(\tilde\nu)=\frac{\tilde\nu^3}{e^{\tilde\nu}-1}
\left\{1+ y \tilde\nu\frac{e^{\tilde\nu}}{e^{\tilde\nu}-1}
\left[\frac{\tilde\nu}{\tanh(\tilde\nu/2)}-4\right]
\right\}.
\end{equation}
Similar distortions due to bulk comptonization are obtained from (\ref{bc2}),

(\ref{bc10})-(\ref{bc13}), for small distortions with $C_f\approx 1$, $\beta_0 \tau\ll 1$,  as
\begin{equation}
\label{bc20}
\delta_{Bulk}=\frac{f_1(\tilde\nu)}{f_0(\tilde\nu)}= \frac{1}{12}\frac{\beta_0\tau}{1+\frac{\beta_0\tau}{4}}\,
 \frac{\tilde{\nu}e^{\tilde\nu}}{e^{\tilde\nu}-1}.
\end{equation}
For the Raleigh-Jeans low frequency side of the spectrum  $h\nu < kT_r$ we have \cite{ZS1}, using  (\ref{bc15}),(\ref{bc18}),

\begin{equation}
\label{bc21}
\delta_{Th}\approx -2y=-\tau_{CL}\, \frac{kT_e}{m_e c^2}, \,\,\, \cite{ZS1} \qquad \delta_{Bulk} \approx
\frac{\beta_0 \tau}{12},
\end{equation}

A thermal comptonization leads to shifting the whole CMB spectrum to higher photon energy region, producing decrease of low-energy photons (see Fig.3). In the case of bulk comptonization the observed photons cross the whole collapsing layer. Half of the matter velocity is directed opposite to the photon trajectory, increasing its frequency , and another half is moving in the same direction with  photons, which are loosing the energy. Actually, the resulting spectrum is moving to the higher frequency also (see Fig.4), but the deviations from the Planck are much smaller. The shift of the spectral maximum in both cases are comparable, but maximum itself is slightly increasing for the bulk comptonization, and is decreasing more for the thermal one.

The distortion in the high energy Wien part of the spectrum due to comptonization is relatively large, and cannot be found correctly in the linear approximation, but its input in the energy is negligibly small. For thermal comptonization in the uniformly expanding universe it is found in \cite{ZS1}. The consideration of this case for a bulk comptonization in a contracting layer is more complicated, and will not be considered here. It seems, that for analysis of observations it is enough to know, that in both cases the high energy part of the spectrum is increasing. It is, probably, not possible to distinguish between two type of comptonization, by observations of high energy part because of substantial indefiniteness of our knowledge about the distribution of parameters of the hot gas in galactic clusters. The observations in the vicinity of maximum of the CMB spectrum could more distinctly indicate the mechanism of comptonization because of different sign of the spectrum distortion in this region.

For qualitative comparison.  energy spectrum distortions by comptonization are presented in Fig.3 for bulk comptonization at $\beta_0\tau=1.2$, $C_f=0.7744$, from (\ref{bc14}), and for a thermal comptonization in Fig.4, for $y=1/60$.

\begin{figure}
\centering
\includegraphics[scale=0.5]{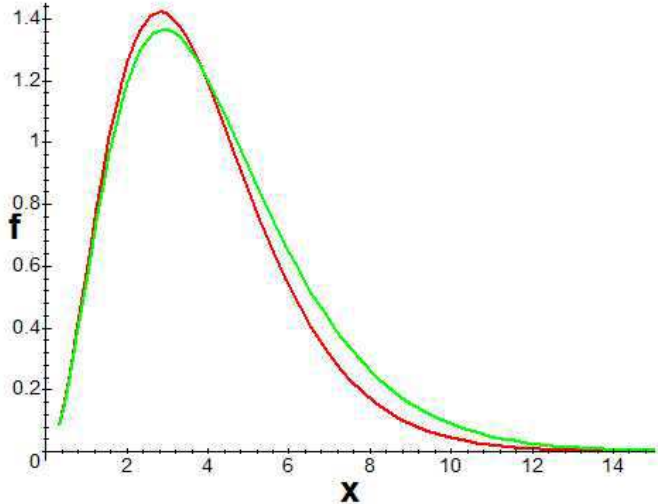}
\caption{Spectral distortion due to comptonization on hot electrons at $y$=1/60, according to \cite{ZS1} and (\ref{bc19})
Plotted are Planck curve (red line), and comptonized curve (green line)}
\label{therm}
\end{figure}

\begin{figure}
\centering
\includegraphics[scale=0.5]{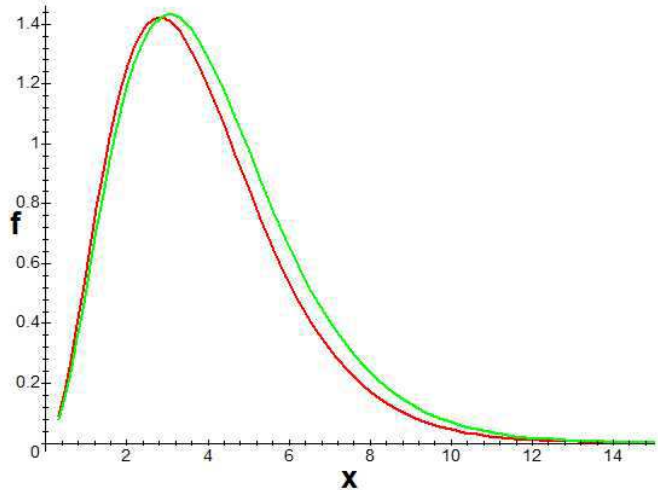}
\caption{Spectral distortion due to bulk comptonization on the contracting layer at $\beta_0\tau=1.2$, 
according to (\ref{bc14}). Plotted are Planck curve (red line), and comptonized curve (green line)}
\label{bulk}
\end{figure}

\section{Conclusions}

If the large scale structure of the Universe was created, even partially, via Zeldovich pancakes, than the fluctuations of the CMB radiation should be formed due to bulk comptonization of black body spectrum on the contracting pancake. Observations indicate \cite{BanNat}, that the secondary ionization was started at early stages, at $z\sim 20$, so that contracting pancakes should have a high enough ionization level for the comptonization onset. It is not possible now to predict certainly the amplitude of these fluctuations, because of uncertainty in the epoch of formation, and masses of the pancakes. Nevertheless, measuring a spectrum of CMB fluctuations could permit to distinguish between thermal and bulk comptonization. In the first the deficit of low energy photons is accompanied by increase of high energy ones, because in the comptonization process the total number of photons is conserved, and decrease of the spectral maximum. In the bulk comptonization on the contracting pancake the excess is obtained also at higher energy side photons, with slight increasing of the maximum, while decrease in the lower part is  less.  May be in the observed spectrum where  both types of comptonization are present: the bulk comptonization at the epoch of pancake formation, and the thermal comptonization at present time on the hot gas in the galaxy cluster.

\section*{Acknowledgements}

The author is grateful to L.G. Titarchuk for useful discussions, and  Ya.S. Lya\-khova for cooperation.
This work was partially supported by RFFI grants No.17-02-00760, 18-02-00619, and RAS Program of basic research 12 "Problems of Origin and Evolution of the Universe".

\end{document}